\documentclass[fleqn,twoside]{article}
\usepackage{espcrc2}

\usepackage{epsfig}

\usepackage{graphicx}
\usepackage[figuresright]{rotating}

\newcommand{\AmS}{{\protect\the\textfont2
  A\kern-.1667em\lower.5ex\hbox{M}\kern-.125emS}}

\title{10 D Euclidean quantum gravity on the lattice.\thanks{ITEP-LAT/2003-30}}

\author{A.I.~Veselov\address[ITEP]{ITEP,
B.Cheremushkinskaya 25, Moscow, 117259, Russia }%
        and M.A.~Zubkov\addressmark[ITEP]\thanks{
        Speaker at the conference. Research supported by the
        grants RFBR 01-02-17456, 03-02-16941, 02-02-17308, RFBR-DFG 03-02-04016,
INTAS 00-00111, CRDF award RP1-2364-MO-02, and by Federal Program of the
Russian Ministry of Industry, Science and Technology No 40.052.1.1.1112. }}

\begin{document}

\begin{abstract}
$10$ D Euclidean quantum gravity is investigated numerically using the
dynamical triangulation approach. It has been found that the behavior of the
model is similar to that of the lower dimensional models. However, it turns out
that there are a few features that are not present in the lower dimensional
models.
%%%\vspace{1pc}
\end{abstract}

% typeset front matter (including abstract)
\maketitle

\section{INTRODUCTION}

Recently it has been paid much attention to the dynamical
triangulation (DT)
 approach to quantization of Riemannian geometry \cite{Triangulation} in
 $4$ dimensions. The choice $D = 4$ is usually motivated as an
 attempt to construct quantum theory of gravity in the real $4$ -
 dimensional world. However, there exist a lot of models, in which
 our world appears as a compactification of higher dimensional
 space (see, for example, \cite{Strings}
  and references therein). Therefore the problem to investigate the
quantization of Riemannian geometry in arbitrary dimension appears. The
correspondent construction is expected to be a part of a hypothetical theory
that would unify all known fundamental interactions.

Until now the dynamical triangulation approach was applied  only
to the cases $D = 2, 3 , 4$, and  $5$. For $D = 2$ the DT model
  has a well - defined
 continuum limit consistent with the predictions of the continuum theory
  \cite{2D}. At $D = 3,4,5$ the Euclidean DT models with the discretized
  Einstein action have two phases: the crumpled phase with infinite fractal dimension
   and the elongated one, which resembles branched polymer model with the
 fractal dimension close to $2$. {\it A priori} it is not clear
 that the behavior of dynamical triangulation model at high $D$ is
 similar to that of lower dimensional models. This motivates our
 choice $D = 10$. We are looking for the qualitative difference
 between the case $D=10$ (which is treated as {\it high} -
 dimensional) and the cases $D=3,4,5$ (that are treated as {\it low} -
 dimensional).

\section{The model.}

In the dynamical triangulation approach the Riemannian manifold
(of Euclidian signature) is approximated by the simplicial complex
obtained by gluing the $D$ - dimensional simplices. The Einstein
action can be expressed through the number of bones and the number
of simplices of the given triangulated manyfold:
\begin{small}
\begin{eqnarray}
&&S  =  - \frac{1}{16 \pi G} \int R(x) \sqrt{g} d^D x\nonumber\\&
=& - \frac{\rm const}{16 \pi G}\sum_{\rm bones} (2\pi - O({\rm
bone}) cos^{-1} (\frac{1}{D}))
\nonumber\\
 & = & - \kappa_{D-2}  N_{\rm bones} + \kappa_{D} N_{\rm
 simplices},\label{S}
\end{eqnarray}
\end{small} where $O({\rm bone})$ is the number of simplices sharing the
given bone, $N_{\rm bones}$ is the total number of bones and
$N_{\rm simplices}$ is the total number of simplices,
$\kappa_{D-2}, \kappa_{D} \sim \frac{1}{G}$.

The functional integral over $D g$ in this approach is changed with the
summation over the different triangulations (we restrict ourselves with the
spherical topology only): $\int D g \rightarrow \sum_{T}$.

We consider the model, in which the fluctuations of the global
invariant $D$ - volume are suppressed. The partition function has
the form: \begin{small}\begin{equation} Z = \sum_T {\rm
exp}(\kappa_{D - 2} N_{D - 2} - \kappa_{D} N_D - \gamma (N_D -
V)^2 )\label{ZVD}
\end{equation}\end{small}
where we denoted $N_D = N_{\rm simplices}$ , $N_{D-2} = N_{\rm
bones}$ , and $\kappa_{D-2} = \frac{\rm const}{8 G}$.
Unfortunately, it is not possible to construct an algorithm that
generates the sum over the triangulations of the same volume. So
the constant $\gamma$ is kept finite. Here
$\kappa_D(V,\kappa_{D-2})$ is already not fixed by the
gravitational constant $G$ and is chosen in such a way that $< N_D
> =  V $. This provides that the volume fluctuates around the required value
$V$.

For the numerical investigations we used Metropolis algorithm in its form
described in \cite{Catterall}. It is based on the so - called $(p,q)$ moves
that are ergodic in the space of combinatorially nonequivalent triangulated
manyfolds \cite{ergodicity}. Due to this property starting from the
triangulation that approximates \cite{Geometry} the given Riemannian manifold
it is possible to reach a triangulation that approximates almost any other
Riemannian manifold. The exceptional cases, in which this is impossible, are
commonly believed not to affect physical results.

\section{Numerical results.}

We made our calculations within the parallel programming environment using the
computation facilities of Joint Supercomputer Center at Moscow (supercomputer
MVS 1000M). The main program was written in $C^{++}$ using modern methods of
object - oriented programming\footnote{We are very obliged to J.Ambjorn and
S.Catterall who have sent us their program codes on the dynamical
triangulations written in Fortran and C respectively.}.

It follows from (\ref{S}) that the Einstein action stimulates the
system to have infinite $O({\rm bone})$ at negative $\kappa_{D-2}$
and minimal $O({\rm bone})$ at positive gravitational coupling.
Therefore, we expect that at finite $V$ and large negative
$\kappa_{D-2}$ the system should exist in the completely collapsed
phase (crumpled phase), in which $O({\rm bone}) \rightarrow {\rm
max}$ \footnote{In 4D this phase is realized in such a way that
almost all the simplices are shared by a certain link.}. At large
positive $\kappa_{D-2}$ we expect the system to exist in the
phase, in which $O({\rm bone}) \rightarrow {\rm min}$. This case
should correspond to the branched polymer system (elongated
phase). The dynamics at intermediate values of $\kappa_{D-2}$ is
{\it a priori} not clear. However, our investigation shows that
similar to the lower dimensional cases the mentioned phases are
the only phases of the $10$ - dimensional system. We investigate
the behavior of $10D$ model for $V = 8000$ and $V = 32000$ . We
considered the values of $\kappa_D$ varying from $-0.1$ to $0.1$.
The phase transition point is at $\kappa_D \sim -0.03 $ at $V =
8000$ and at $\kappa_D \sim -0.01 $ at $V = 32000$.

One of the most informative characteristics of the triangulated manyfold is the
average volume ${\cal V}(R)$ of the balls
 (of the
constant radius $R$)\footnote{The ball of radius $R$ is the set of
simplices $v$ such that the geodesic distance between their
centers and the center of the given simplex $s$ is $\rho(v,s) \leq
R$.} as a function of this radius. At some value of $R$ ${\cal
V}(R)$ becomes constant. This value of $R$ is the averaged largest
distance between two simplices of the manyfold, which is also
called the diameter $d$. We found that for $V = 8000$ at $1 < R <
\frac{d}{2}$ the dependence of ${\rm log} \,{\cal V}$ on ${\rm
log} \,R$ is linear. For $V = 32000$ the same takes place at $4 <
R < \frac{d}{2}$. The slopes give us the definition of the fractal
(Hausdorff) dimension of the manifold: ${\rm log} \,{\cal V} =
{\rm const} + {\cal D}{\rm log} \,R$.

\begin{figure}[htb]
%%%\vspace*{-0.75cm}
\epsfig{file=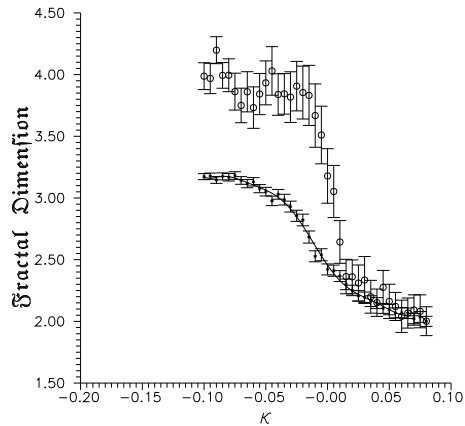,width=6.5cm,angle=0} \vspace*{-0.75cm}
\caption{\it The fractal dimension. Data for $V=8000$ is
represented by
 the points and the solid
 line. Data for $V=32000$ is represented by the circles.}
\label{fig.04} \vspace*{-0.4cm}
\end{figure}

 The mean fractal dimension ${\cal D}(V)$  of the manyfold for
the observed volumes $V$ as a function of $\kappa_{D-2}$ is
represented on the Fig.~\ref{fig.04}. The figure indicates that
there is a phase transition at critical $\kappa_{D-2} =
\kappa_c(V)$. For the observed volumes $\kappa_c$ appears to be
small negative value ($\kappa_c(8000) \sim -0.3$ and
$\kappa_c(32000) \sim -0.1$). For $\kappa > \kappa_c$ the fractal
dimension is close to two in accordance with the supposition that
 the correspondent phase
resembles branched polymers. For $\kappa < \kappa_c$ ${\cal
D}(8000) \sim 3.2$ and ${\cal D}(32000) \sim 4$. This is in
accordance with the expectations that this phase has a singular
nature and corresponds to ${\cal D}(\infty) = \infty$.

\begin{figure}
 \epsfig{figure=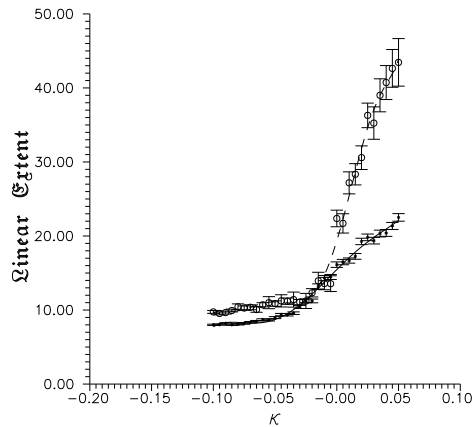,width=6.5cm,angle=0}
 \caption{\label{fig.01} \it The  linear extent.  Data for $V=8000$ is represented by
 the points and the solid
 line. Data for $V=32000$ is represented by the circles and the dashed line.
 \label{fig.1}}
\end{figure}

The linear size of the system can be evaluated as $V^{\frac{1}{\rm
D}}$. So, we expect that in the crumpled phase the diameter   $d
\sim 10$ while in the elongated phase   $d \sim 100$. The direct
measurements of the diameter as well as another parameter called
linear extent (see \cite{Catterall}) confirm these
expectations\footnote{ $D = 10$ should be treated as bare
dimension while the fractal dimension $\cal D$ is treated as
"physical". Therefore to investigate the system of linear size $L$
we need $V = L^{\cal D}$ instead of $V = L^{10}$.}. The linear
extent ${\cal L}$ is defined as the average distance between two
simplices of the triangulation:
\begin{equation}
{\cal L} = \frac{1}{V^2}\sum_{u,v}<\rho(u,v)>
\end{equation}
Due to its construction $\cal L$ should be close to half a diameter.  Our
results on $\cal L$ are represented in the figure Fig.~\ref{fig.01}. One can
see that the linear size of the manyfold is increasing very fast in the
elongated phase, while in the crumpled phase it remains almost constant (and
close to that of $3$, $4$ and $5$ dimensional models \cite{Triangulation}). We
found that the fluctuations of the linear extent are almost absent in the
crumpled phase and are of the order of the mean size of the manyfold in the
elongated phase.

\section{Conclusions and discussion.}

We found that in general the behavior of the $10$ - dimensional model is
similar to that of $3,4$ and $5$ - dimensional models. However, there is the
qualitative difference between the entropy properties of the dynamical
triangulations in these cases. If we turn off the action ($\kappa_{D-2} = 0$)
only the entropy is responsible for the dynamics. At $V \leq 32000$ and
$\kappa_{D-2} = 0$ the ten - dimensional system behaves like a branched polymer
system while at $D = 3,4,5$ the system with turned off action becomes
completely collapsed (crumpled phase).

\end{document}